\documentclass[10pt]{article}
\usepackage[BOE]{express}
\usepackage{changepage}
\usepackage{xcolor}
\usepackage{subfig}
\usepackage{pgf}

\begin{document}
\bibliographystyle{osajnl} 

\title{Multi-purpose SLM-light-sheet microscope}

\author{Chiara Garbellotto, Jonathan M. Taylor$^*$}

\address{University of Glasgow, University Avenue, Glasgow, G12 8QQ, UK}


\email{\authormark{*}jonathan.taylor@glasgow.ac.uk}



\begin{abstract}
By integrating a phase-only Spatial Light Modulator (SLM) into the illumination arm of a cylindrical-lens-based Selective Plane Illumination Microscope (SPIM), we have created a versatile system able to deliver high quality images by operating in a wide variety of different imaging modalities. When placed in a Fourier plane, the SLM permits modulation of the microscope's light-sheet to implement imaging techniques such as structured illumination, tiling, pivoting, autofocusing and pencil beam scanning. Previous publications on dedicated microscope setups have shown how these techniques can deliver improved image quality by rejecting out-of-focus light (structured illumination and pencil beam scanning), reducing shadowing (light-sheet pivoting), and obtaining a more uniform illumination by moving the highest-resolution region of the light-sheet across the imaging Field of View (tiling). Our SLM-SPIM configuration is easy to build and use, and has been designed to allow all of these techniques to be employed on one optical setup compatible with the OpenSPIM design. It also offers the possibility to choose between three different light-sheets, in thickness and height, which can be selected according to the characteristics of the sample and the imaging technique to be applied.
We demonstrate the flexibility and performance of the system with results obtained by applying a variety of different imaging techniques on samples of fluorescent beads, Zebrafish embryos, and optically cleared whole mouse brain samples. Thus our approach allows easy implementation of advanced imaging techniques while retaining the simplicity of a cylindrical-lens-based light-sheet microscope.
\end{abstract}



\section{Introduction}

Selective Plane Illumination Microscopy (SPIM)\cite{{Huisken2004}} is becoming increasingly popular for live  fluorescence imaging in developmental biology. 
Compared to confocal microscopy, which also performs optical sectioning, the use of a static light-sheet to illuminate only the imaging focal plane means that SPIM offers important advantages including rapid snapshot acquisition and reduced photobleaching of the sample \cite{Huisken2016}. Snapshot acquisition is particularly important for capturing the 3D structure of rapidly changing organisms or scenes; photobleaching is minimized by the confinement of the illumination light to a single plane, thus ensuring that the only fluorophores excited are those that lie in the plane currently being imaged.
Despite these attributes making SPIM particularly well-suited for \emph{in vivo} imaging, a classical SPIM implementation suffers from a number of issues including:
\begin{itemize}
\item{Shadow artefacts}: parts of the sample will absorb or scatter the side-launched light-sheet, generating dark stripes behind them, elongated parallel to the illumination direction. 
\item{Scattering}: when illuminating a  plane inside the sample, light emitted by the excited fluorophores has to travel through sample tissue in order to be collected by the imaging objective, which means it inevitably undergoes some scattering on the imaging path. This results in undesirable out-of-focus background in the images, leading to reduced image contrast. Tissue scattering also affects the propagation of the light-sheet itself, resulting in even more out-of-focus light, in this case coming from out-of-focus fluorophores excited by the scattered light-sheet.
\item{Limited field of view}: even in the absence of scattering, the illumination delivered by a Gaussian light-sheet is not uniform in thickness across the image Field of View (FoV). A Gaussian beam generates a light-sheet with a certain waist size (thickness of the light-sheet at its focus) and extent (Rayleigh length). This shape of the light-sheet results in an uneven illumination across the image FoV, with better optical sectioning around the beam waist, where the light-sheet is at its thinnest, and poorer optical sectioning (more out-of-focus excitation) at the sides of the image, generated by the thicker parts of the sheet.
 \end{itemize}
 A variety of modifications to SPIM light-sheet microscopy have been proposed to tackle some of these issues, and similarly for the closely-related technique of DSLM light-sheet microscopy (Digital Scanned Laser Light-sheet Fluorescence Microscopy \cite{Keller2008}),
where a light-sheet is synthesized by rapid scanning of a focused 2D Gaussian beam. Shadows can be reduced by illuminating the sample from multiple directions \cite{Huisken2007,Itoh2016}, and using Bessel \cite{Fahrbach2012,Planchon2011,Olarte2012} or Airy \cite{Vettenburg2014} beams instead of Gaussian beams permits a more uniform illumination across a larger FoV. One way to reduce the effects of scattering from tissue surrounding the imaged plane is to use a DSLM configuration (where a synthetic light-sheet is formed from a scanned Gaussian beam) in conjunction with a rolling confocal slit on the detection camera to reject scattered light \cite{Baumgart2012}. Methods based on structured illumination, such as HiLo \cite{Mertz2010} and the method proposed in \cite{Breuninger2007} which we refer to as the 3-phase method, can also help enhance image contrast by reducing the out-of-focus contribution. The technique known as tiling can be used to extend the limited FoV over which high-quality depth sectioning can be achieved using a simple Gaussian beam \cite{Gao2015}. In this case each plane in the sample is imaged multiple times, each time with the light-sheet focused at a different lateral position in the FoV. The final image of the plane is then created by stitching together adjacent vertical stripes taken from the different images, each stripe containing only the part of the image generated by the thinnest part of the light-sheet.
 
The aim of the present paper is to present a single SPIM system that uses a spatial light modulator (SLM) to create a simple and flexible platform for performing the aforementioned advanced light-sheet microscopy techniques,
as well as permitting automatic adaptive light-sheet positioning for best focus (autofocusing). Our setup consists of a basic SPIM microscope (illumination arm with cylindrical lens, imaging chamber with sample holder and translation stage, imaging arm)  with the addition of a phase-modulating SLM in the illumination arm. With the SLM conjugated to a Fourier plane in the optical path of the illumination beam (SLM illuminated by a collimated light beam), we demonstrate that it is possible to modulate and move the resultant light-sheet to implement a wide range of imaging techniques without having to move any mechanical part in the microscope. While SLMs have previously been used for certain specific purposes in light-sheet illumination ~\cite{Vettenburg2014,Wilding2016,TomsThesis,Li2014}, we will show that a single microscope design is well-suited for applying a wide variety of different optical techniques on the same imaging platform.

This work is organized as follows. In the next section we present the optical design of our SLM-SPIM microscope, discuss the motivation behind our design choices and give details about the different samples used in our experiments and the way they were mounted for imaging. In Section \ref{Section3} we introduce some of the imaging techniques that can be performed with our system, and include experimental results obtained imaging fluorescent beads in agarose, Zebrafish embryos, and cleared mouse brain samples. All the raw data and the MATLAB codes used to produce figures, plots and calculations presented in Section \ref{Section3} can be found in the data repository \cite{{SLM_SPIM_glasgow_repository}}.

\section{SLM-SPIM microscope} 
\label{SLM-SPIM microscope}

SLMs are versatile pixellated phase (or amplitude) modulation devices that have already found many applications in the field of optical microscopy \cite{Maurer2011},  where they have been used to control the sample illumination \cite{Quirin2013} or as a Fourier mask in the imaging path \cite{Lee2013,Zammit2014}. SLMs have previously been described by a few authors in SPIM systems, both in the illumination and in the imaging path (for example to deliver structured illumination \cite{Li2014} or to correct for aberrations \cite{Wilding2016}). Recent work in parallel with our own has presented the SSPIM (Structured SPIM \cite{Aakhte2018}), a DSLM microscope with an incorporated SLM which offers the option to shape the beam profile (Gaussian, Airy, Bessel, Lattice) as well as performing tiling and structured illumination. In contrast, our work showcases the use of a programmable SLM device in a more classical scan-free, cylindrical-lens-based SPIM such as that used for the OpenSPIM design \cite{Pitrone2013}.

In our SPIM we use a reflective phase-only liquid crystal SLM, placed in a Fourier plane in the optical path of the illumination beam. The optical setup is illustrated in Figure \ref{scheme setup}, while details of the components and devices used are summarized in Table \ref{setup table}. The laser beam coming out of the optical fiber is collimated, expanded (to fill the size of the SLM active surface), and directed onto the SLM, which modulates the phase of the beam's wavefronts. Two lenses and a cylindrical lens are then used to create the light-sheet and conjugate the plane of the SLM to the back focal plane of the illumination objective. This conjugation means that what the objective focuses (in one direction) on the sample is the far field diffraction pattern corresponding to the phase-shift pattern displayed on the SLM, i.e. its Fourier transform. A second system optimized for imaging cleared mouse organs (glycerol/CLARITY) shares common laser launch optics and SLM with our water-immersion system, but has a separate final part of the illumination arm (last spherical lens, cylindrical lens and launching objective), glycerol chamber, and a vertically mounted imaging arm (components included in Table \ref{setup table}). The patterns displayed on the SLM can easily be rotated and customized for either one of the systems, permitting us to perform the different imaging techniques on both water- and glycerol-immersed samples.

As illustrated in Figure \ref{scheme setup}, our water-immersion system also allows easy switching between three different optical configurations on the laser launch path. Each of these interchangeable modules has a different positioning of the cylindrical lens and final spherical lens. This allows us to choose between three light-sheets of different thickness and height, as well as offering different conjugations between the SLM and the objective lens (discussed in detail below). Experimental profiling of the light-sheet (using a small fluorescent bead translated through the light-sheet) has confirmed a FWHM (at sheet waist) of  around $ 2     \, \mathrm{\mu}$m for setup 1, $  3 \, \mathrm{\mu}$m for setup 2 and $5 \,$ $\mathrm{\mu}$m  for setup 3. The sheet height is $\sim$4 mm for setup 1, $\sim$2 mm for setup 2, and $\sim$0.6 mm for setup 3. 
%
%
\subsection{Design considerations}

In the light-sheet launch path, it is necessary to place the cylindrical lens after the SLM. Otherwise, if the SLM is appropriately conjugated to a pupil or image plane, a line focus will be formed on the SLM and no meaningful phase control would be available along one axis. However, it follows from this choice that it is not possible for the SLM to be simultaneously conjugate to the pupil plane in both the horizontal and vertical axes simultaneously (and similarly, neither can it be simultaneously conjugate to the object plane in both axes).  These considerations mean that care is required in designing the light-sheet launch optics to ensure that the SLM provides the necessary degrees of freedom to manipulate the light-sheet as required for an experiment. 

\begin{figure}
\begin{adjustwidth}{-12em}{-12em}
\centering\includegraphics[width=13cm]{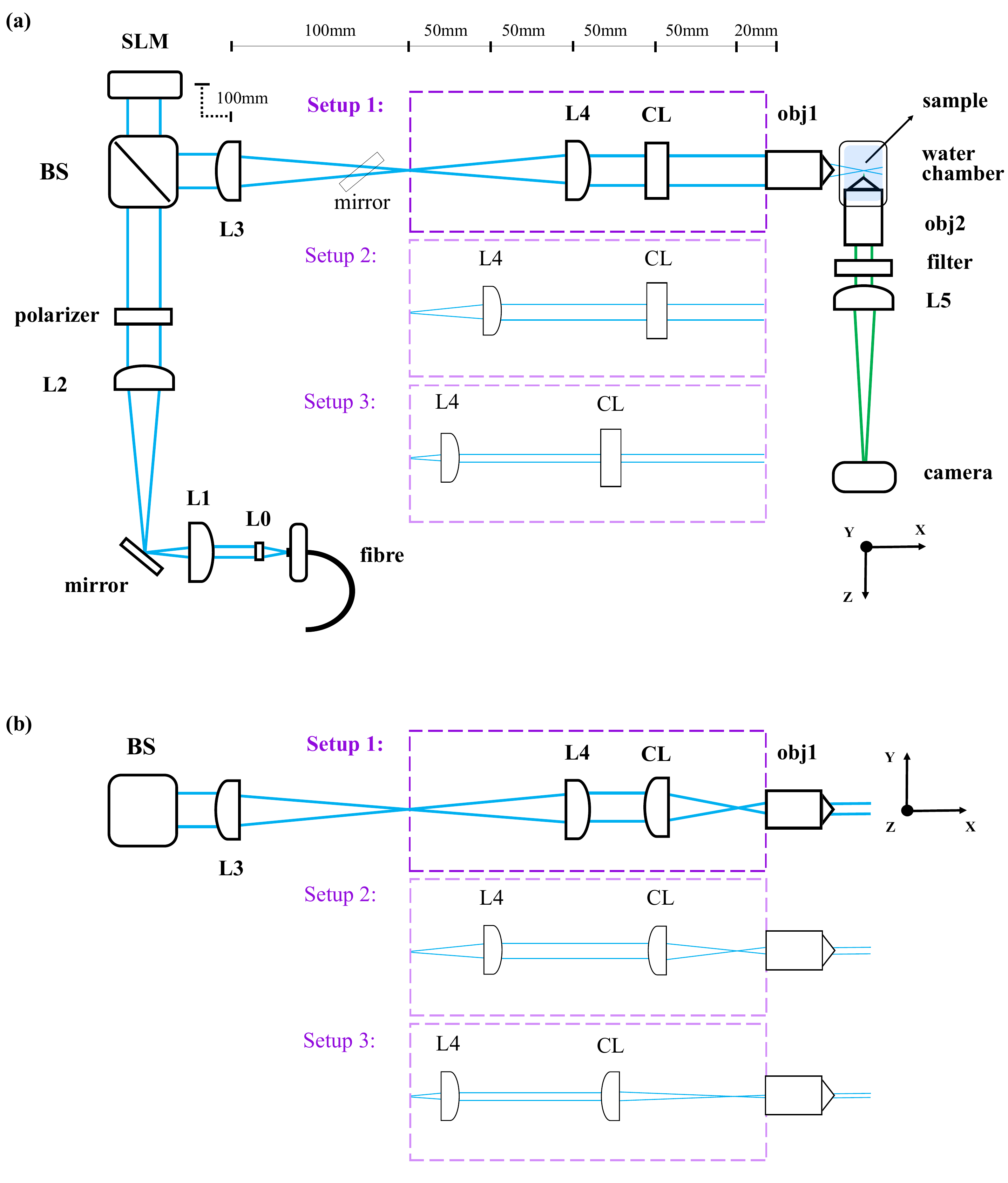}
\end{adjustwidth}
\caption{Optical scheme of our SLM-SPIM, with a top view of the system in (a) and a side view of its launching arm in (b) (see Table \ref{setup table} for details of the individual components).  Changing the position of the last two lenses before the illumination objective allows us to switch between three different setups, yielding different sheet heights and thicknesses, and changing the conjugation of the SLM with the sample plane. (a) View of the SLM-SPIM from above. The cylindrical lens has no optical power in this plane. A mirror placed before the SLM (bottom right corner) permits adjustment of the vertical position of the light-sheet in the sample plane. A second mirror can easily be inserted after L3 and used to redirect the laser beam to the side (upwards in this figure), onto a second illumination arm, (not included in this scheme) ending in the glycerol chamber. The glycerol illumination arm consists of (see Table \ref{setup table} for details): L3 (shared with the water-imaging system) $\leftarrow$160mm$\rightarrow$ CL2 $\leftarrow$40mm$\rightarrow$ L6 $\leftarrow$100mm$\rightarrow$ obj3 $\rightarrow$ glycerol chamber. The glycerol imaging arm is mounted vertically above the glycerol  chamber and is composed of a glycerol dipping objective (obj4), a tube lens (L7) and the same camera used for the images in water. (b) Side view of the  final part of the SLM-SPIM illumination arm, with three different possible configurations.}
\label{scheme setup}
\end{figure}

The three interchangeable lens modules mentioned above offer different conjugations between the plane of the SLM and the focal plane of the illumination objective, each of which is optimal for different families of applied beam shaping techniques.
In a top-down view of our system design (Figure \ref{scheme setup}a), it can be seen that setups 1 conjugates the SLM with the back focal plane of the illumination objective: the two lenses L3 and L4 are separated by the sum of their focal lengths, the SLM is at $f_3$ from L3 and the back focal plane of the objective is at $f_4$ from L4. 
\begin{table}
\begin{center}
\begin{tabular}{r p{9.3cm}}
\textbf{Lenses} & \textbf{L0:} achromatic doublet (Thorlabs, AC064-013-A-ML), $f=13$ mm; \textbf{L1:} plano-convex, $f=35$ mm; \textbf{L2:} plano-convex, $f=100$ mm; \textbf{L3:} plano-convex, $f=100$ mm;  \textbf{L4:} plano-convex, $f=100$ mm for setup1, $f=50$ mm for setup 2 and $f=25.4$ mm for setup 3; \textbf{CL:} cylindrical lens, $f=50$ mm for setup 1 and 2, $f=80$ mm for setup 3; \textbf{L5:} plano-convex, $f=100$ mm;\\ 
 & \\
\textbf{Other optical elements} & \textbf{BS:} beam splitter; \textbf{polarizer}: linear polarizer;\textbf{filter:} Green Fluorescent Protein (GFP) filter (central wavelength 525 nm); \textbf{obj1:} 10$\times$ Nikon Plan Fluorite Imaging Objective, 0.3 NA, 16 mm WD (Working Distance); \textbf{obj2:} 40$\times$ Nikon CFI APO NIR Objective, 0.80 NA, 3.5mm WD;\\ 
 & \\
\textbf{SLM} &  Hamamatsu LCOS-SLM (Liquid Crystal on Silicon Spatial Light Modulator) serie X13138;\\ 
\textbf{SLM head} & \textbf{pixels:} 1272 $\times$ 1024; \textbf{pixel size:} 12.5 $\mathrm{\mu}$m; \textbf{effective area size:} 15.9 mm $\times$ 12.8 mm; \textbf{fill factor:} 96 \%; \\ 
\textbf{SLM controller} & \textbf{input signal:} DVI-D; \textbf{DVI signal format (pixels):} 1280 $\times$ 1024; \textbf{input signal levels:} 256; \textbf{DVI used frame rate:} 60Hz; \\ 
 & \\
\textbf{Laser} & OBIS coherent laser; \textbf{wavelength:} 488 nm; \\ 
 & \\
\textbf{Camera} & XIMEA MD028xU-SY; \textbf{sensor active area:} 8.8 mm $\times$ 6.6 mm; \textbf{resolution:} 1934 $\times$ 1456, 2.8 Mp; \textbf{pixel size:} 4.54 $\mathrm{\mu}$m; \textbf{frame rate:} 56.9 fps; \textbf{dynamic range:} 71.1 dB;\\ 
&\\
\textbf{Glycerol system} &  Illumination arm, after the SLM, L3 and the pop-in mirror: \textbf{CL2:} cylindrical lens, $f=60$ mm; \textbf{L6:} plano-convex, $f=100$ mm;  \textbf{obj3:} 5$\times$ ZEISS EC Plan-Neofluar Objective, 0.16 NA, 18.5mm WD. Imaging arm: \textbf{L7:} plano-convex, $f=150$ mm; \textbf{obj4:} 20$\times$ ZEISS Clr Plan-Neofluar Objective, 1.0 NA, 5.6mm WD;\\ 
 & \\
\end{tabular}
\end{center}
\vspace*{-6mm}
\caption{List of components used (with reference to Figure \ref{scheme setup}).}
\label{setup table}
\vspace*{-4mm}
\end{table}
However, the optical power of the cylindrical lens means that, when viewing the system from the side (Figure \ref{scheme setup}b), the SLM is not conjugate to either a pupil or image plane. In contrast, setup 2 gives the opposite situation and, in its side view, it conjugates the plane of the SLM with the focal plane of the launching objective: L4 and the cylindrical lens are separated by the sum of their focal lengths and the back focal plane of the objective is at $f_{cl}$ from the cylindrical lens. Setup 3 gives a conjugation similar to the one of setup 2, but was designed to give a good compromise between a perfect conjugation (achievable with the use of a cylindrical lens with $f_{cl} = $ 75 mm, instead of the $f_{cl} = $ 80 mm we decided to use for this setup) and the combination of high flexibility in tilting the light-sheet and a high demagnification of the incoming beam (both of which are achieved using a longer $f_{cl}$), resulting in a light-sheet with thicker waist and longer Rayleigh length (see below for a more detailed discussion about these choices).

Most of the imaging techniques we performed on our system can be achieved using any of the three setups, but because of the different sheet height, sheet thickness and SLM conjugation they provide, different setups would be the preferred choice for different imaging techniques. Setup 1 gives the thinnest light-sheet at beam waist, and is therefore the best one to use for experiments such as the light-sheet tiling ones (Section \ref{tiling section}), where the only part of the light-sheet which is actually used to create the final image is its central, thin waist. The conjugation of the SLM with the focal plane of the launching objective generally makes setup 2 the most appropriate for shadow suppression experiments (Section \ref{section pivoting}): the light-sheet is in this case tilted in the sample plane, and the perfect conjugation of the SLM with the center of the FoV assures that the light-sheet does not shift vertically while being tilted (which would result in a change in image brightness dependent on the light-sheet tilt). Setup 3 gives a thicker light-sheet, which on the other hand also means a more even illumination across the FoV (longer Rayleigh length). This makes it a good choice for experiments such as the pencil beam scanning in Section~\ref{section pencil beam scanning}, where the light-sheet illumination is substituted with a vertically scanned focused beam. The use of a thicker but more uniform beam also helps reduce the time needed to generate a homogeneously illuminated image of the entire FoV. 

The choice of what setup to use for a particular experiment should also depend on the characteristic of the sample to be imaged.  For the experiments which we could perform with more than one of the three setups, we chose to use Setup 1, which is the one that we find gives the best light-sheet for imaging fluorescent beads and Zebrafish embryos, with a vertically uniform illumination across the used FoV and a $2 \, \mathrm{\mu}$m thick light-sheet waist. In Section \ref{section pivoting} (shadow suppression experiments) we give an example of a situation in which the type of sample strongly influences the setup choice, further demonstrating the advantages of working with an easily reconfigurable system.

In our experiments, we use the SLM in a refractive mode (applying blazed gratings).
This means that most of the light modulated by the SLM is concentrated in the first diffracted order, but some power is always lost in higher orders and in the so-called 0th order (containing the specularly-reflected light that is not modulated by the SLM). In order to eliminate the 0th order we apply a further, constant phase ramp to the SLM, to displace the desired first order from the 0th order. The SLM is tilted such that the first order lies on the optical axis, and the 0th (and higher) orders can then be blocked by a mask placed in the first focal plane after the SLM. When we apply different phase patterns to the SLM, the trajectory and structure of the first order beam is altered slightly, but it is still allowed to pass by the mask.

\subsection{Samples}
\label{samples}
For our experiments with the water objective, the samples where mounted in a length of FEP tubing (Fluorinated Ethylene Propylene, 1.3 mm ID $\times$ 1.6mm OD, Adtech Polymer Engineering Ltd) attached to the end of a  syringe. For calibration and alignment purposes, and for the tiling experiments, the FEP tube was filled with beads (polystyrene beads, 0.2 $\mu$m Dragon Green, Bangs Laboratories Inc) in a 1.5 \% low melting point agarose solution (Agarose, High-EEO/Protein Electrophoresis Grade, Fisher Scientific). When imaging ex-vivo Zebrafish embryos, the fish was placed in a FEP tube filled with system water. 
For our experiments we used the Zebrafish nacre mutant (reduced pigment), with the cardiac muscle labelled with green fluorescent protein (\textit{myl7:eGFP}). 
The fish specimen was preserved in formalin (10\% Formalin solution, neutral buffered, Sigma-Aldrich). For the experiments on cleared mouse samples, we used a mouse brain prepared with the CLARITY method \cite{Chung2017}. The sample was placed in a quartz cuvette (UV fused quartz glass, 
Thorlabs part CV10Q3500F) filled with 85\% glycerol,  which was sealed while excluding any air bubble from inside. The cuvette was then immersed in the chamber filled with 85\% glycerol, and held horizontally underneath the dipping imaging objective.

\section{Results and analysis}
\label{Section3}
\subsection{Structured Illumination}
\label{Section SI}
\begin{figure}[h!]
\vspace*{-5mm}
\begin{adjustwidth}{-12em}{-12em}
\centering
\includegraphics[width = 0.61\linewidth]{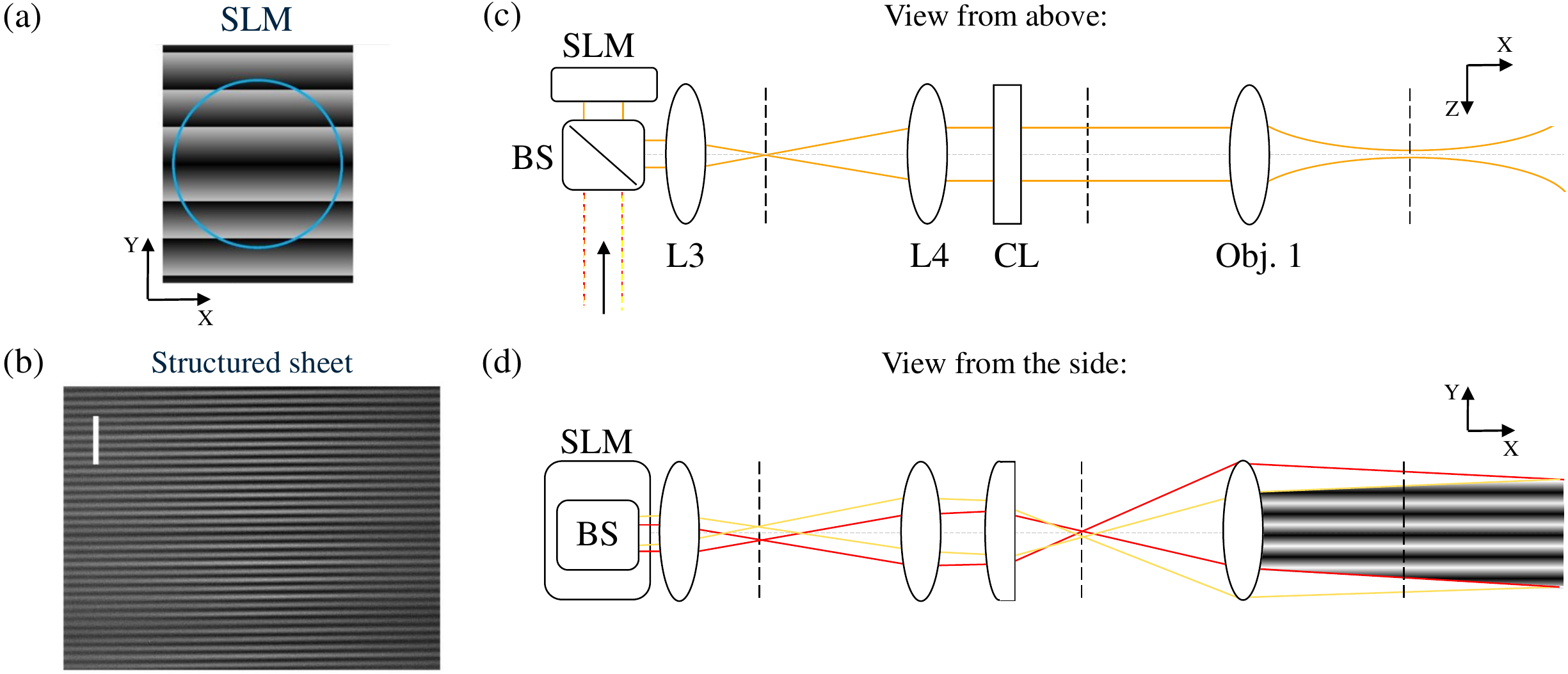}
\end{adjustwidth}
\vspace*{-2mm}
\caption{(a) Pattern displayed on the active area of the SLM to create a sinusoidally modulated light-sheet; the blue circle indicates the footprint of the collimated input beam. (b) Experimental image of a modulated light-sheet obtained with our method, imaged in aqueous Fluorescein dye diluted in water, revealing an interference pattern with a period of $\sim \!\! 10 \mathrm{\mu}$m (white to white);  scale bar: 50 $\mathrm{\mu}$m. (c) Schematic showing the optical path (distances and sizes not to scale) of the light reflected off the SLM. The two beams follow the same optical path when viewed in the \textit{xz} plane, forming two light-sheets on the same plane.  (d) In the \textit{xy} plane, the two sheets propagate at different angles, generating an interference pattern in the sample plane.}
\label{SI_SLM}
\end{figure}

Two mutually-coherent light-sheets that propagate in the same plane ($xy$ plane), but at a different angle with respect to the optical axis of the illumination arm, will interfere in the sample plane and generate a light-sheet with sinusoidal modulation along \textit{y }(Figure \ref{SI_SLM}b). This type of intensity-modulated light-sheet can be used to perform two different structured illumination techniques: HiLo \cite{Mertz2010, Schroter2012}, and the 3-phase method \cite{Neil1997, Neil1998}. In the 3-phase  method, three images are taken by translating the same sinusoidal light-sheet by precise spatial phase shifts of: 0, 2/3$\pi$ and 4/3$\pi$. The three images $\mathrm{I}_{1,2,3}$ are then combined using the following formula:
\begin{equation}
\vspace*{-1.5mm}
\label{formulaSI}
\mathrm{I} = \frac{\sqrt[]{2}}{3}\Big[(\mathrm{I}_1-\mathrm{I}_2)^2 +(\mathrm{I}_1 - \mathrm{I}_3)^2 + (\mathrm{I}_2 - \mathrm{I}_3 )^2 \Big] ^{1/2},
\end{equation}
to yield a final image $\mathrm{I}$ with reduced out-of-focus background and hence enhanced contrast. This 3-phase technique has been implemented on a digitally-scanned light-sheet microscope by fast time-modulation of the scanned beam \cite{Keller2010}, and in 2007 on a SPIM microscope using a grid-projection approach \cite{Breuninger2007}. HiLo imaging has also been implemented previously on a digitally-scanned light-sheet microscope; it requires only two raw images, one acquired with a normal, uniform light-sheet and one with a modulated light-sheet~\cite{Mertz2010}.

\begin{figure}[h!]
\begin{adjustwidth}{-12em}{-12em}
\centering
\includegraphics[width = 13.4cm]{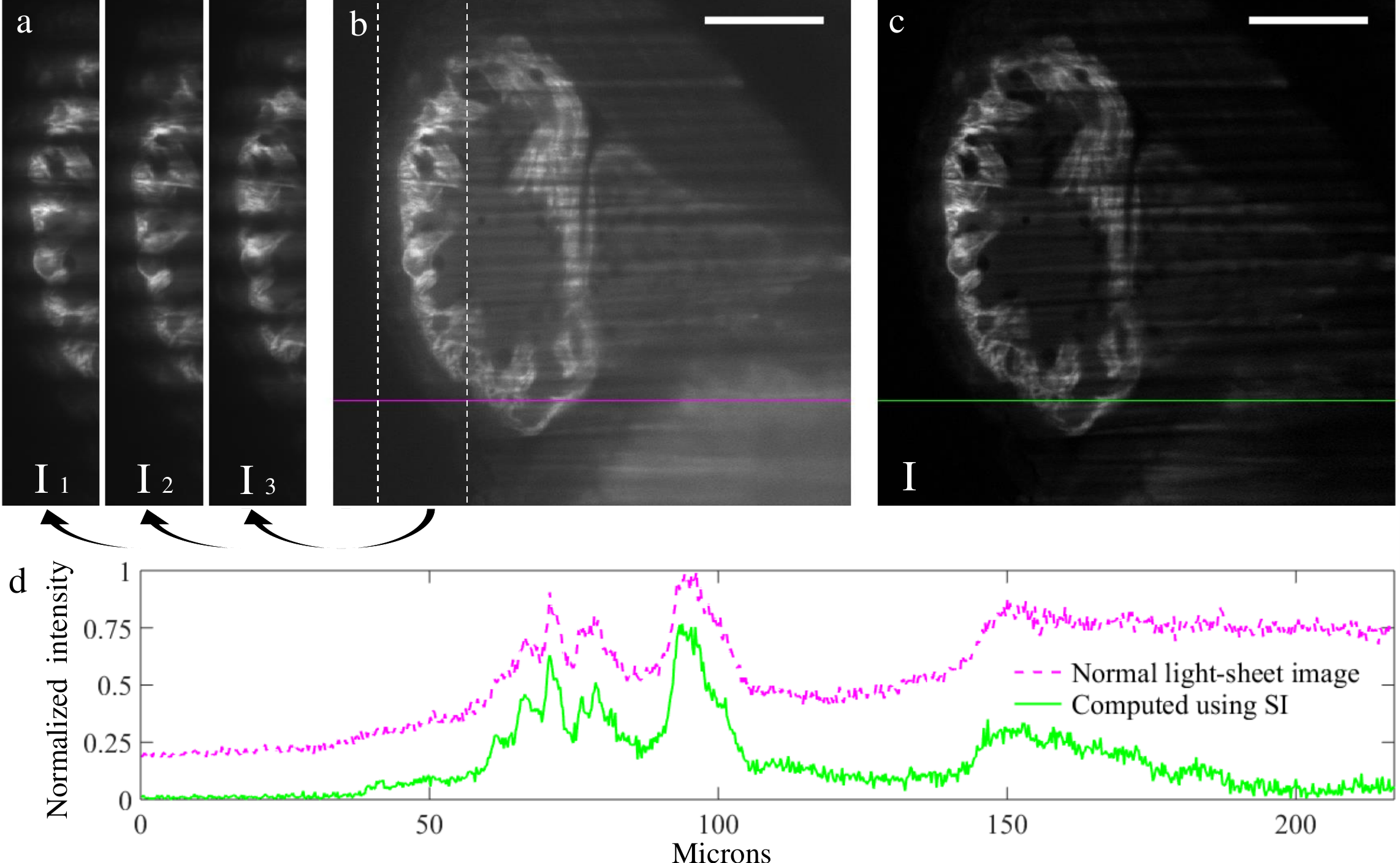} 
\end{adjustwidth}
\vspace*{-2mm}
\caption{3-phase structured illumination performed using setup 1. (a) Cropped views of the three images, $I_{1,2,3}$, taken with a modulated light-sheet (period = 20 $\mathrm{\mu}$m on the sample plane). The sinusoidal pattern for the second and third images is shifted by a phase of $\frac{2}{3}\pi$ and $\frac{4}{3}\pi$ with respect to the first image. (b) Image acquired with a normal, non-modulated light-sheet. (c) Image obtained by combining $I_{1}$, $I_{2}$ and $I_{3}$  using Equation~\ref{formulaSI}. (d) Intensity profile along the same row in images (b) and (c), to visualize the achieved background reduction and improved image contrast (values normalized to the global maximum of the two plotted lines). Scale bars: 50 $\mathrm{\mu}$m.}
\label{SI_results}
\vspace*{-5mm}
\end{figure}

To generate the two interfering sheets, we display two opposite sawtooth patterns simultaneously on the top and bottom halves of the SLM, as in Figure \ref{SI_SLM}a. As the beam diffracts off the SLM it is thereby split into two half-beams, propagating in the image plane ($xy$ plane) but at two opposite angles $\alpha$ and $-\alpha$ from the optical axis of the launching arm. When the two half-beams meet again and interfere in the sample plane, they generate the desired sinusoidal pattern (Figure \ref{SI_SLM}b). The period of the final illumination pattern is defined by the propagation angle, which is in turn determined by the sawtooth period on the SLM. In order to shift the illumination pattern by a desired phase (to perform the 3-phase method), the equivalent optical phase shift can simply be added to one of the two halves of the SLM. Figure \ref{SI_results} shows results from the application of this structured illumination technique to image the heart of a formalin-fixed Zebrafish embryo specimen (4 days post-fertilization) expressing GFP fluorescence.
 
To quantify the improvement in image contrast obtained as a result of the out-of-focus background reduction, we calculated the standard deviation of the energy-normalized histograms~\cite{Keller2010} of the structured illumination (Figure \ref{SI_results}c) and the normal light-sheet images (Figure \ref{SI_results}b). This standard deviation can be calculated using the following formula:
\begin{equation}
\vspace*{-2mm}
\label{formulaSTD}
\sigma_\mathrm{N} = \sqrt[]{\frac{\displaystyle\sum_{i} \bigg( \frac{\mathrm{I}_i - \bar{\mathrm{I}}}{\sum\mathrm{I}_i} \bigg)^2}{C - 1}},
\end{equation}
where $C$ is the total number of pixels in the image, $i$ ranges from 1 to $C$, $\mathrm{I}_i$ is the intensity value of the $i$-th pixel, $\sum\mathrm{I}_i$ is the sum of all the pixels values in the image, and $\bar{\mathrm{I}} = \sum\mathrm{I}_i/C$ is the mean intensity value of the image. As more extensively explained in \cite{Keller2010}, the ratio between the $\sigma_\mathrm{N}$ values of two images can be used to quantify the change in image contrast, with an higher $\sigma_\mathrm{N}$ value corresponding to a better image contrast. For the images shown in Figure \ref{SI_results}b and \ref{SI_results}c we calculated this ratio to be  $\sigma_\mathrm{N}(c)/\sigma_\mathrm{N}(b) \sim 2.6$.

It should be noted that our approach to generate the two interfering light-sheets cannot be used with setup 2 and 3. In fact, in the case of perfect conjugation between the SLM and the waist of the light-sheet (or almost perfect conjugation with setup 3) the desired interference appears only on one side of the imaging FoV. Instead, the mis-conjugation offered by the side view of setup 1 moves the edge of the interference region to the side, allowing the two half-beams to interfere across the whole FoV. A possible future alternative could be to develop an approach analogous to the one used in \cite{Judkewitz2014}. The SLM would be divided into vertical stripes instead of into two halves, and the two blazed gratings displayed on alternate stripes. This method would allow structured illumination experiments to be performed with all our three setups, but it would give rise to extra diffraction orders which would reduce efficiency, increase out-of-focus excitation and bleaching, and might therefore require masking out.

\subsection{Tiling}
\label{tiling section}
\begin{figure}[h!]
\begin{adjustwidth}{-12em}{-12em}
\centering\includegraphics[width=13.5cm]{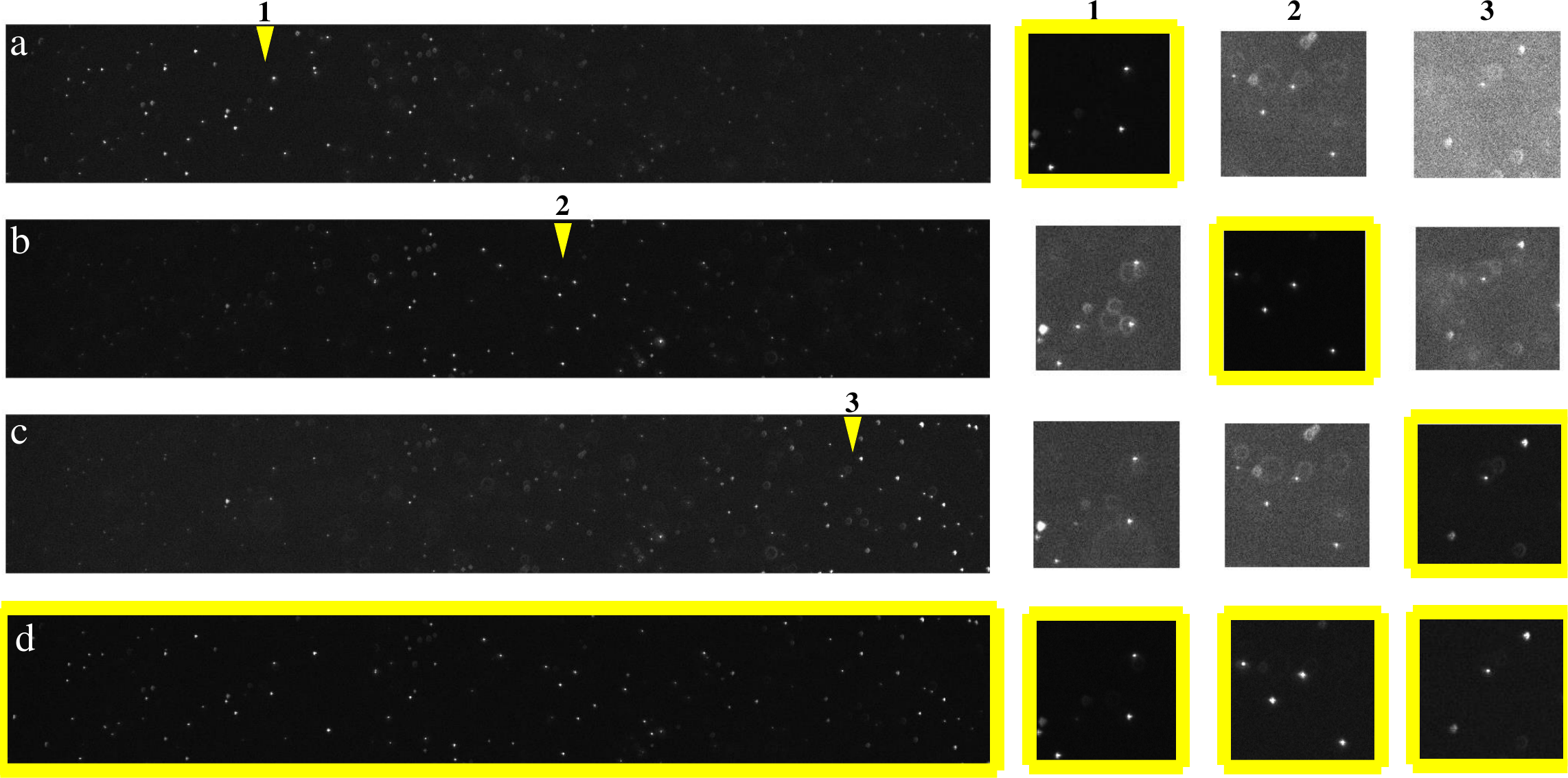}
\end{adjustwidth}
 \vspace*{-2mm}
\caption{Tiling technique demonstrated with 0.2 $\mathrm{\mu}$m fluorescent beads. We acquired eight images of the same plane of beads, each image with the light-sheet focused at a different position (laterally) in the FoV. Left: (a,b,c) Same horizontal stripe taken from the second, fifth and eighth of the eight images taken (only three images shown for sake of clarity): in each image the position of the sheet waist  (indicated by the yellow arrowheads) can be recognized by the brightness of the beads and the reduced number of out-of-focus beads. (d) Image obtained tiling the eight images, i.e. retaining only the sharpest vertical stripe taken from each of them. Right: zooms on beads taken from three different lateral positions (1,2,3) in the images, with each row showing how the same beads appear in the corresponding image to the left. Each of these zoomed images has been normalized to its own maximum value for clarity. Notice how, looking at one column at a time (i.e. the same sets of beads), as the position of the sheet's waist gets further away from the position of the beads, the relative amount of light illuminating out-of-focus features increases. The composite tiled image (d) gives the best optical sectioning (most of the light concentrated on in-focus features) throughout the entire FoV.}
\label{Tiling results}
 \vspace*{-6mm}
\end{figure}

Tiling is a technique proposed to work around the trade-off between light-sheet thickness and length \cite{Gao2015, Fu2016}. The ideal light-sheet would stay thin across the whole imaging FoV. Light-sheets created by a focused Gaussian beam only remain thin over the beam's Rayleigh length, and thus thinner sheets diffract and spread more rapidly as they propagate. To mimic the illumination delivered by an ideal thin and long light-sheet, one can use the waist of a short, thin sheet to image only one part of the FoV, then move the sheet's waist laterally and build up, step by step, an image of the entire FoV. In our case, the light-sheet can be focused at different distances from the launching objective  by displaying a Fresnel lens pattern on the SLM. Images acquired by illuminating the same plane inside the sample but changing the focus of the light-sheet (along its propagation direction, i.e. in the $x$ direction in the imaging coordinate system) are then combined to obtain a final image. Of each of the initial images, only a restricted vertical stripe generated by the thin part of the light-sheet is allowed to contribute to the final image. Results obtained using this technique imaging fluorescent beads can be seen in Figure \ref{Tiling results}. For the tiling experiments we chose to use setup 1, which is the one that gives the thinnest light-sheet (best $z$ resolution, but smallest effective FoV in $x$).

\subsection{Multi-angle illumination for shadow suppression}
\label{section pivoting}

\begin{figure}[t]
\begin{adjustwidth}{-12em}{-12em}
\centering\includegraphics[width=13.5cm]{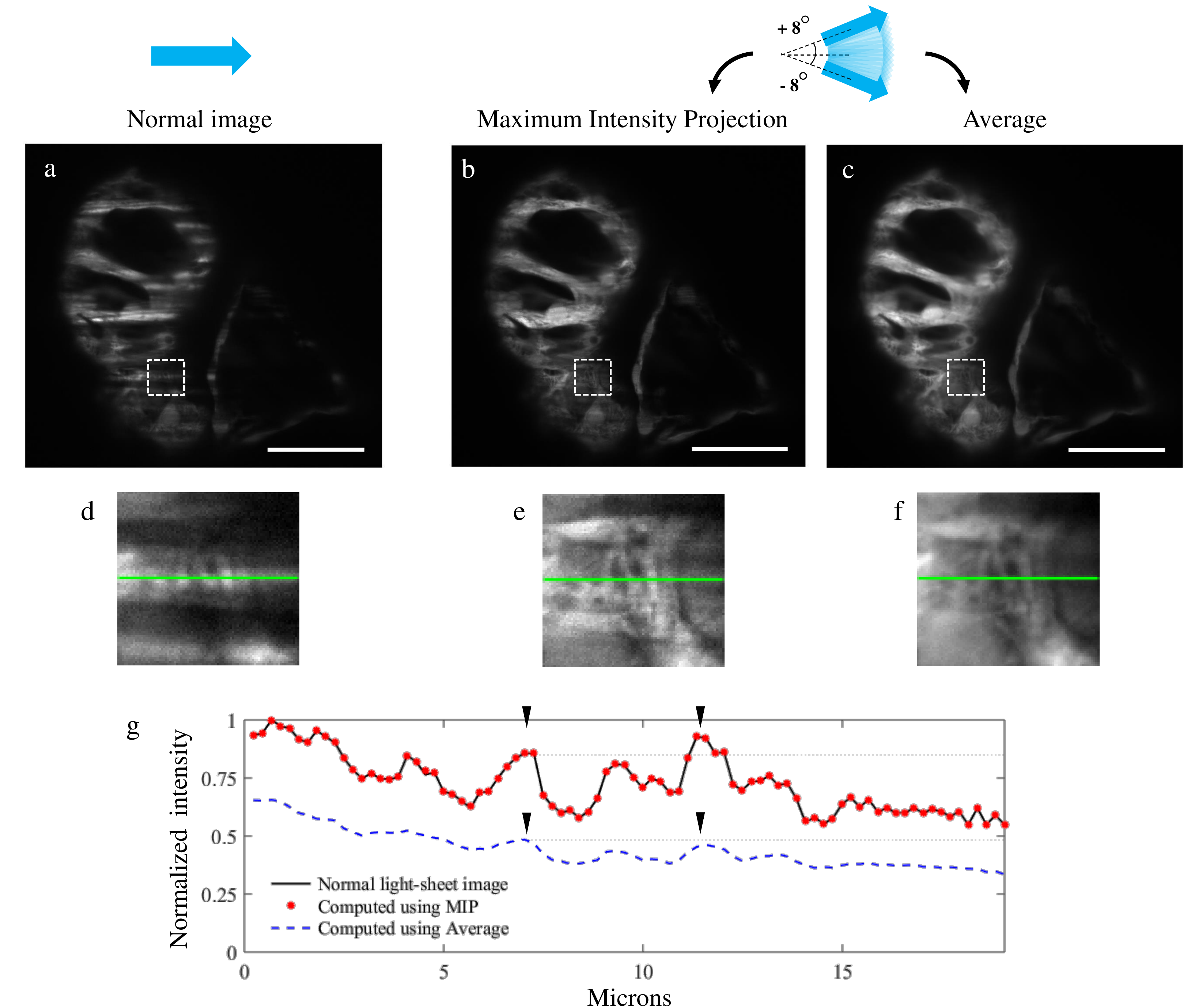}
\end{adjustwidth}
\caption{Shadow suppression using the light-sheet pivoting technique. (a) Image of a formalin-preserved Zebrafish embryo heart (4~dpf) acquired with a normal light-sheet, using setup 3. (b) Image obtained by computing the Maximum Intensity Projection (MIP) of a stack of seventeen images, acquired with the light-sheet propagating at different angles, equally spaced within $\pm8$ degrees. (c) Image obtained by averaging the same seventeen images used for (b). (d-f) Zoomed-in views of the dashed line boxes in images (a-c). Each of these images has been normalized to its own maximum value. (g) Intensity profile of the same horizontal line in images (d-f). This plot shows how the MIP allows to preserve the original image contrast and, with respect to averaging, a more accurate representation of the true intensity profile: notice how averaging (blue dashed line) distorts the relative intensity of the two peaks indicated by the black arrowheads, making the left peak appear as brighter than the one on the right. Intensity values in these plots are normalized to the global maximum of the three plotted lines. Scale bars: 50 $\mathrm{\mu}$m.}
\label{pivoting}
\end{figure}

To be imaged using SPIM, a sample first of all needs to be transparent enough for the light-sheet to propagate through it. Parts of the sample that strongly scatter or absorb the excitation light create a visible shadow behind them in the ``downstream'' direction of beam propagation (in our case to their right, since the light-sheet illuminates the sample from the left). This shadow effect can be reduced by combining  illumination generated by light-sheets propagating at different angles \cite {Huisken2007, Itoh2016}. In both these previous works the illumination angle was modulated at high speed ($>1$ kHz) to provide a range of illumination angles within a single image exposure. In our system (using a slower SLM) light-sheets with different propagation directions can be created by displaying different sawtooth patterns on the SLM. A shadow-free image can be obtained by  switching between the different light-sheets within the exposure time of a single image, or by recording one image for each light-sheet inclination and combining the different images afterwards. This second approach, despite being less efficient in terms of image acquisition/computing time, offers the flexibility to permit what we propose as an alternative and improved algorithm for combining the acquired images to obtain shadow suppression: Maximum Intensity Projection (MIP). When the light-sheet is tilted through a range of different angles within a single image exposure, the resulting shadow-free image is generated from the \emph{sum} of all the fluorescence excited by each light-sheet inclination. This technique can be replicated by computing the sum of a set of images acquired each with the light-sheet propagating at a different inclination. An alternative way of combining these images is to compute the Maximum Intensity Projection (MIP) of the whole stack: for each pixel compare the values assigned to that pixel on each image and only keep the maximum one. This results in a final image where each pixel takes on the value from the raw image where that region was experiencing minimal shadowing. As can be seen in Figure \ref{pivoting}, the image obtained using MIP not only preserves a better image contrast when compared to the one obtained by averaging, a but it also assures a more accurate representation of the true intensity profile across the image. In fact, computing the average of a set of images acquired with the light-sheet propagating at different angles results in an alteration of the true image intensity profile: parts of the sample which are well-illuminated in all the images (i.e. are not affected by shadows) are inherently seen as brighter than those that are only illuminated in some of the images. Using MIP on the other hand ensures that the final intensity of each part of the sample only depends on the intensity observed when it is illuminated without obstruction, and not on the number of images which \textit{agree }with that value. In Figure \ref{pivoting} we compare a normal light-sheet image and the two alternative shadow suppression algorithms, averaging and MIP. A zoom-in on a region strongly affected by shadows shows the shadow suppression results obtained by combining seventeen images acquired with the light-sheet propagating at different angles, equally spaced within $\pm8$ degrees. Before being combined either with MIP or averaging, the seventeen images were properly rescaled to account for the diffraction efficiency of the SLM, which changes with the angle of propagation of the diffracted first order, with a higher tilt corresponding to a dimmer light-sheet. We selected an horizontal line across a region of the sample which is not affected by any shadows in the normal light-sheet image (green line in Figure \ref{pivoting}(d)), and compared its intensity profile with the intensity profile of this same line in the two images computed for shadow suppression. This plot helps viualising the reduced image contrast offered by the averaging technique, and also its unpredictable distortion of the  original intensity profile. 

Because of the conjugation of the SLM plane with the center of the FoV, setup 2 is the most appropriate setup to be used for the shadow suppression experiments, provided it can deliver a high enough tilt angle with high diffraction efficiency. As the SLM is used to send the first diffraction order to different directions, it generates light-sheets that propagate at different angles but overlap entirely again on each plane conjugate to the plane of the SLM, as they do on the SLM itself. In the case of setup 2, as the light-sheet is tilted using the SLM, its rotation in the sample plane  happens around the center of the FoV, which is in fact conjugated to the plane of SLM.
In setup 1 the situation is different: the rotation of the light-sheet happens around a position that is to the side of the FoV, such that a tilt also corresponds to an undesired vertical shift of the light-sheet in the images. As the tilt angle increases, the bright central part of the Gaussian sheet shifts away (vertically) from the FoV, which is illuminated by less and less light.\\
One other thing to keep in mind is that for shadow suppression the best results are obtained by using a large range of angles for the incoming light-sheets. In our system, a limit to the maximum achievable tilt angle is set by the SLM pixel spacing. 
By considering the magnification of the relay optics within our microscope,
we can find the relation between the tilt angle of the light-sheet in the sample plane and the angle at which it originally propagates as it is diffracted off the SLM:
\begin{equation}
\label{formula angles}
\mathrm{sin}(\theta_2) = \frac{f_1 f_3}{f_2 f_4} \, \mathrm{sin}(\theta),
\end{equation}
where $\theta$ is the propagation angle after the SLM, $\theta_2$ is the propagation angle in the sample plane, and $f_1$, $f_2$, $f_3$ and $f_4$ are the focal lengths of the 
four lenses through which the beam passes
before reaching the sample plane (see Figure \ref{scheme setup}b), with $f_4$ being the focal length associated with the launch objective. Finally, we can approximate $\mathrm{sin}(\theta)$ with $\lambda$/$p$ and obtain:
\begin{equation}
\label{formula angles lambda/p}
\mathrm{sin}(\theta_2) = \frac{f_1 f_3}{f_2 f_4} \, \frac{\lambda}{p},
\end{equation}
where $\lambda$ is the wavelength (in our case 488 nm) and $p$ is the pixel period of the sawtooth pattern displayed on the SLM: the smaller the period of the sawtooth pattern on the SLM, the larger the angle at which the light-sheet generated by the first diffracted order propagates. On the other hand, decreasing the sawtooth pattern period size also reduces the number of SLM pixels used for each period, and this coarser approximation of the ideal pattern results in a less efficient concentration of the diffracted light in the first diffracted order, i.e. the resultant light-sheet is dimmer. Even for larger pixel periods, this varying diffraction efficiency means that larger-angle sheets will be somewhat dimmer, and prior to further processing we rescale each image to compensate for the different brightnesses of the light-sheets they have been generated with.

In order to maintain a sufficiently bright first order, we required a minimum period of four SLM pixels.  Using Equation ~\ref{formula angles lambda/p}  we can verify that a sawtooth period of four SLM pixels corresponds, with setup 2, to a tilt angle of $\sim 3^\circ$. For our experiments on the embryonic Zebrafish heart, we found that this maximum tilt angle was insufficient for good shadow suppression. We therefore decided to perform these experiments with setup 3, which gives adequate conjugation between the SLM plane and the center of the imaging FoV but offers a much bigger range of possible tilt angles, with a 50 $\mu$m period (four SLM pixels) corresponding to a tilt angle of $\sim 8.8^\circ$. Figure \ref{pivoting} illustrates the results obtained combining seventeen images acquired with setup 3, with the light-sheet propagating at different angles, equally spaced within $\pm8^\circ$, imaging the heart of an ex-vivo 4 dpf Zebrafish embryo.
 
\subsection{Pencil beam scanning (synthetic DSLM)}
\label{section pencil beam scanning}
Part of the fluorescence excited in the illuminated plane undergoes scattering by the intervening tissue before reaching the detector, resulting in an increased diffused background signal and a loss of contrast. In DSLM light-sheet systems, where the light-sheet is formed by a rapidly-scanned 2D Gaussian beam (a ``pencil'' beam), the beam scan can be combined with a confocal rolling shutter on the camera to suppress the background signal~\cite{Baumgart2012}. e can replicate this same approach by recording a sequence of full-frame images as the pencil beam is scanned across the FoV, and then create the final image by applying a synthetic confocal slit to each raw image (i.e. masking out all rows except those where the pencil beam should have appeared). This post-acquisition confocal slit method is implemented for example in~\cite{Fahrbach2012}, where each image is multiplied by a smooth Gaussian mask centered and aligned with the illumination beam.  In order for the mask to correctly select the desired part of each image,  the position and inclination of the illumination beam  needs to be known, either from previous calibration or from the acquired images. In fact, both implementations of the confocal slit technique (rolling shutter, and post-acquisition masking) require precise alignment and size-matching between the illumination beam and the detection line used. 

Our system enables us to implement the post-acquisition confocal slit method, and we here propose an alternative way to process the raw images and achieve background rejection, obtaining a technique that combines the ease of a standard, full-frame acquisition with a simple and calibration-free postprocessing procedure. We scan the illumination beam across the entire FoV, acquiring a single full image for each position of the beam. We then combine these images into a 3D stack of size $n \times m \times i$, where $n \times m$ is the image size in pixels and $i$ is the number of images taken. The final contrast-enhanced image can be obtained by simply computing the Maximum Intensity Projection of the stack of images along its third dimension (dimension of size $i$).

To explain and justify this procedure, let us concentrate on how the intensity value of a single pixel changes as we scan through the stack of images. Let ($x$,$y$) be the position of the pixel in the image, and ($x_s$,$y_s$) its corresponding location in the sample plane,  and assume we expect to detect some non-scattered signal from ($x_s$,$y_s$). The value of pixel ($x$,$y$) will be very low in the images taken with the pencil beam positioned far away from ($x_s$,$y_s$), somewhat higher as the beam gets closer to it and more scattered light reaches pixel ($x$,$y$), and it will be at its highest when ($x_s$,$y_s$)  is directly in the path of the incoming beam. The Maximum Intensity Projection therefore gives, for each pixel ($x$,$y$), the intensity received by it in the image recorded with the illumination beam giving best overlap with ($x_s$,$y_s$), the position in the sample that maps onto that pixel.  And this is exactly what we wish to retain in our reconstructed image.
%
%
%
%
%
%

Acquiring entire images but then only retaining some pixel rows from each image is of course a relatively slow and inefficient way of implementing the pencil beam scanning technique.
\begin{figure}[b]
\vspace*{-4mm}
\begin{adjustwidth}{-12em}{-12em}
\centering\includegraphics[width=13.2cm]{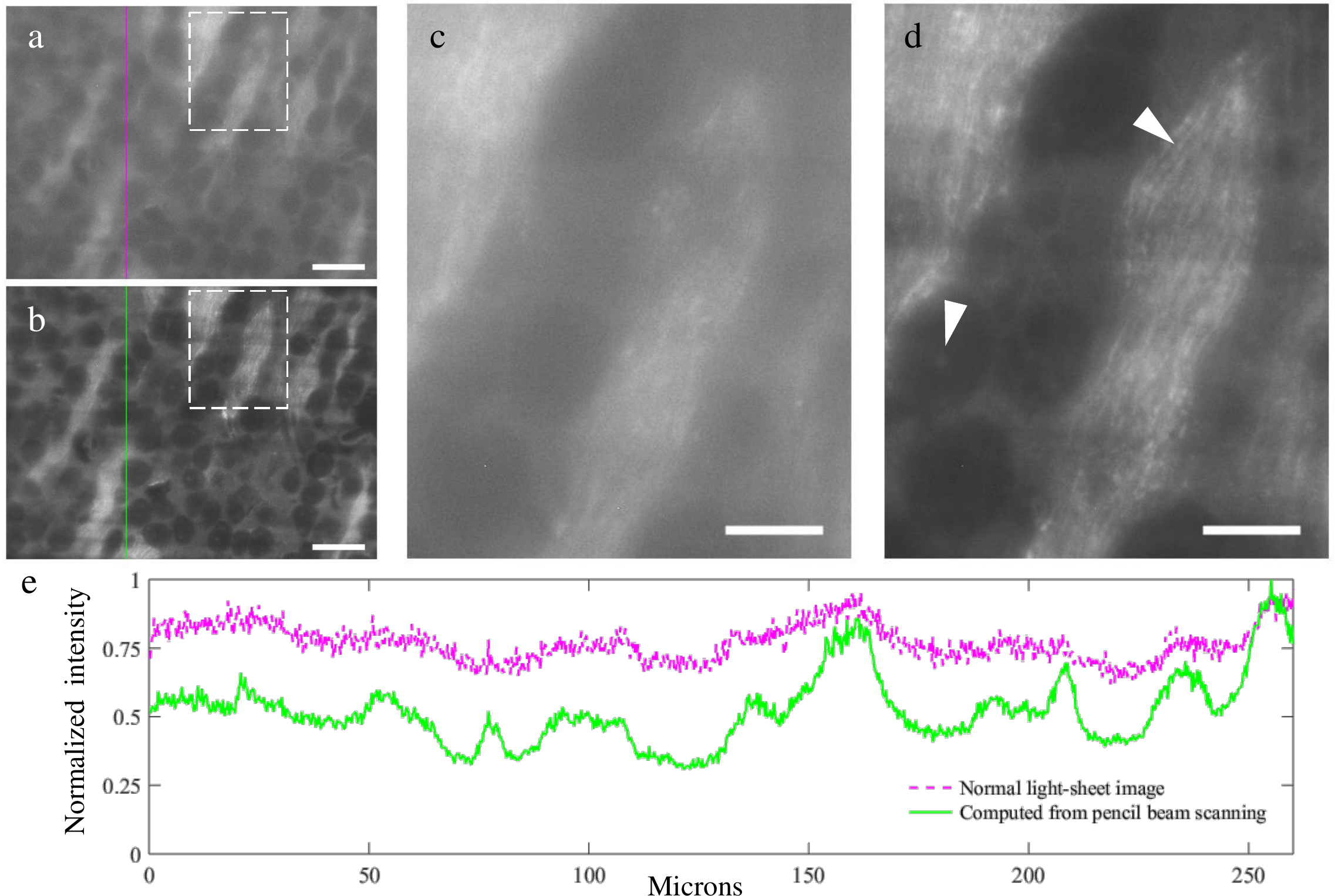}
\end{adjustwidth}
\vspace*{-1mm}
\caption{Pencil beam scanning technique applied on a cleared whole mouse brain sample, imaged in glycerol. (a) Image acquired with a normal light-sheet. (b) Image generated with the scanning pencil beam technique, using our reconstruction procedure (based on a Maximum Intensity Projection) on a set of 200 raw images, each taken with the horizontal pencil beam focused at a different height in the sample plane. Scale bars: 50 $\mathrm{\mu}$m. (c,d) Zoomed-in views of the dashed line rectangles in images (a,b), to highlight some of the faint features (left arrowhead) and fine structures (right arrowhead) revealed by the pencil beam scanning technique. Scale bars are here of 20 $\mathrm{\mu}$m. (e) Cross section of the normalized intensity along the same column in images (a) and (b) (values normalized to the global maximum of the two plotted lines).}
\label{PencilBeam_MouseBrain}
\end{figure} However, this shows how our flexible imaging system can easily be used to explore the feasibility of new imaging modalities, obtaining good pilot results without investing the time and effort needed to build and calibrate a high-speed dedicated rolling-confocal-slit DSLM system. Our post-processing approach based on a simple Maximum Intensity Projection also has the benefit of not requiring any calibration of the pencil beam position and orientation on each image plane.

In our implementation the sample is illuminated by a regular 2D-focused beam, which we generate by displaying an opposing cylindrical lens phase function on the SLM (note that the same could be achieved by physical removal of the cylindrical lens from the system). The beam is scanned using the SLM in order to sequentially illuminate, line by line, the entire in-focus plane.
The phase function applied to our SLM is the sum of the cylindrical lens pattern and a linear phase ramp. In practice the combination of these two functions yields a cylindrical phase function that translates vertically  on the SLM as the pencil beam is scanned (or laterally on the SLM, in case of our experiments with the glycerol setup). 

The improved contrast achievable using this technique is particularly valuable when imaging deep in highly scattering samples. To demonstrate this we performed the pencil beam scanning technique on cleared whole mouse brain samples, with results shown in Figure \ref{PencilBeam_MouseBrain}. For this experiments we programmed the SLM to make the pencil beam translate with steps of  $ 1.4 \, \mathrm{\mu}$m in the sample plane (corresponding to $\sim$6 pixels in the image, with the pencil beam having a FWHM of $\sim 25$ pixels). A total of 200 images were taken to cover the region of interest shown in Figure \ref{PencilBeam_MouseBrain} (1146$\times$1556 pixels).

\subsection{Autofocusing}
By displaying a horizontal phase ramp on the SLM, the light-sheet can be moved in \textit{z}, i.e. towards and away from the imaging objective. This offers a natural method for optimizing the position of the light-sheet to coincide with the focal plane of the camera, without having to move the imaging objective or the tube lens. Particularly for high-numerical-aperture imaging in thick samples, this optimization must often be performed on a per-sample basis, and even when moving to a different location in the sample.

\begin{figure}[h]
\begin{adjustwidth}{-12em}{-12em}
\centering\includegraphics[width=14cm]{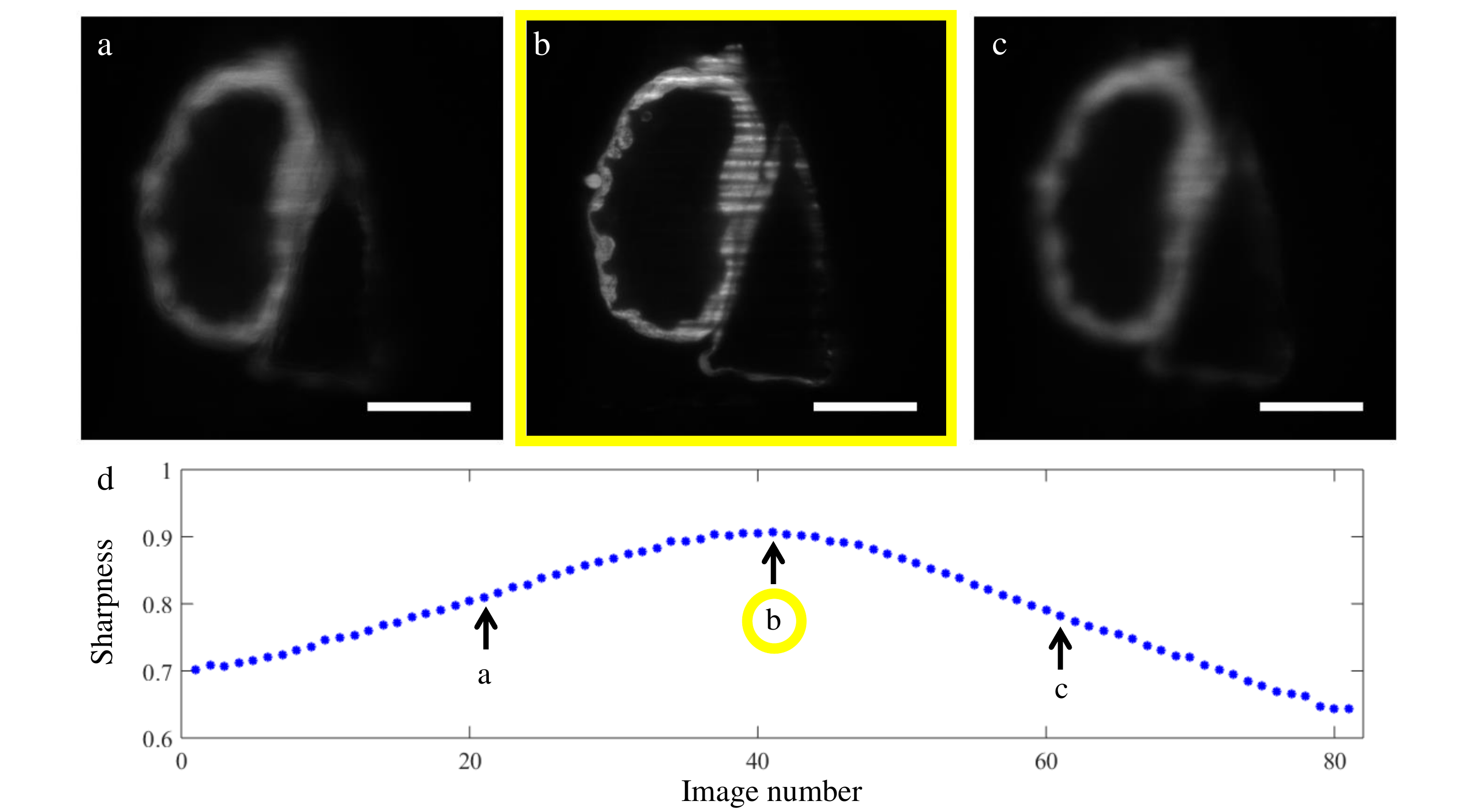}
\end{adjustwidth}
\caption{Autofocusing experiment using sharpness metric of Equation \ref{Eq. Sharpness} on an ex-vivo 4 dpf Zebrafish embryo's heart (using setup 1). In this illustration, eighty-one images were taken using the SLM to move the sheet to different positions with respect to the imaging objective (with steps  of 0.5 $\mathrm{\mu}$m, for a total range of  40 $\mathrm{\mu}$m). (a) and (c) show images taken with the light-sheet in an out-of-focus plane, while (b) is the image that the sharpness measurements identified as the one with best focus (i.e. light-sheet at the correct distance from the imaging objective). (d) Sharpness values, one for each image, with highest value indicating the plane of best focus. Note that in practice a smaller number of images would be taken, and the sharpness interpolated using an appropriate function, to quickly find the optimum focus. Scale bars: 50 $\mathrm{\mu}$m.}
\label{Autofocusing}
\end{figure}

We developed a MATLAB script to automatically optimize the light-sheet position to the plane of best focus. First, the light-sheet is scanned over a certain range in $z$ (chosen by the user) around its rest position (flat SLM), recording one image for each position of the light-sheet. The images are analyzed and the light-sheet is  moved to the position that yielded the image with best focus. To evaluate the quality of the focus of each image, we use the sharpness metric proposed in \cite{Walker2009} and used for adaptive optics on a SPIM in \cite{Bourgenot2012}. This metric quantifies the image focus through a measure of the ratio between the high and low spatial frequency content of the image, and is defined as follows:
\begin{equation}
\label{Eq. Sharpness}
\mathrm{S} = \frac{\sum\limits_{\mathrm{N}_\mathrm{p}}^{}\mid \mathcal{F}[\mathrm{I}(x,y)]\mid_{\mathrm{masked}}}{\sum\limits_{\mathrm{N}_\mathrm{p}}^{}\mid \mathcal{F}[\mathrm{I}(x,y)]\mid_{\mathrm{unmasked}}},
\end{equation}
where $\mathrm{I}(x,y)$ is the intensity of pixel (\textit{x},\textit{y}), $\mathrm{N}_\mathrm{p}$ is the number of pixels in the image, and $\mathcal{F}$ denotes the Fourier transform. A rectangular mask is applied to  the  2D power spectral density (PSD) of the original image ($\mathcal{F}[\mathrm{I}(x,y)]$), in order to mask out its central values,  representing the lowest spatial frequencies contained in the image. The sharpness value S is then given by the sum of the absolute values of $\mathcal{F}[\mathrm{I}]_{\mathrm{masked}}$ divided by sum of the absolute values of $\mathcal{F}[\mathrm{I}]_{\mathrm{unmasked}}$  ($0 < \mathrm{S} < 1$).  
As the images become more blurred (moving away from the plane of best focus), their low spatial frequency content increases with respect to the high frequency content, which means that the mask that subtracts the lowest frequencies has a stronger effect on $\mathcal{F}[\mathrm{I}]$, resulting in a lower value for S. The maximum S value identifies the image with best focus (see sharpness plot in Figure \ref{Autofocusing}).\\ In the experiment presented in Figure \ref{Autofocusing} the sharpness metric was calculated using a  band-pass mask  with a cut-on spatial frequency (in the sample plane) of 0.009 $\mathrm{\mu m}^{-1}$(11$\times$9 pixels) and a cut-off spatial frequency of 0.2275 $\mathrm{\mu m}^{-1}$ (201$\times$151 pixels). 

\section{Conclusion}
We have shown how a phase-only liquid crystal SLM can be introduced as a simple modification to the illumination arm of a SPIM, to give  a flexible, versatile system able to deliver high quality images by applying a range of advanced light-sheet imaging techniques. Imaging fluorescent beads,  Zebrafish embryos and optically cleared whole mouse brain samples, we have demonstrated how the SLM-SPIM can be used to apply: structured illumination and  pencil beam scanning techniques to reduce the out-of-focus content of the images; light-sheet pivoting to reduce the effect of shadows; light-sheet tiling to obtain a more uniform illumination across the image FoV and improve optical sectioning; automated focus optimization. Our modular system also gives the option to choose between three different light-sheets, allowing to select the sheet's thickness and height according to the characteristic of the sample and the imaging technique to be performed. We have also proposed new, computationally-undemanding image reconstruction methods based on the maximum intensity projection operation.

With its simple, functional design and the use of a computer-reconfigurable SLM, we believe our system represents an ideal platform for manipulating the illuminating light-sheet to apply a range of advanced imaging techniques on a single SPIM microscope, and also to explore  combinations of multiple techniques and potentially trial new ones.

\section*{Acknowledgments}

Chiara Garbellotto is supported by a studentship from the EPSRC CDT in Intelligent Sensing and Measurement, Grant Number EP/L016753/1. We thank Andrew Tobin, Sophie Bradley and team (Glasgow University) for the CLARITY-treated mouse brain samples, and Martin Denvir, Carl Tucker and team (Edinburgh University) for the zebrafish samples.

\section*{Declaration}
The Authors declare no conflict of interest.

\end{document}